\pdfoutput=1

\documentclass[11pt]{article}

\usepackage[final]{acl}

\usepackage{amsmath} 
\usepackage{times}
\usepackage{courier}
\usepackage[framemethod=TikZ]{mdframed}
\usepackage{latexsym}
\usepackage{multirow}
\usepackage{multicol}
\usepackage{booktabs}
\usepackage[normalem]{ulem}
\usepackage{adjustbox}
\usepackage{graphicx}
\usepackage[scaled]{helvet}
\usepackage{geometry}
\usepackage{arydshln}

\usepackage[T1]{fontenc}
\usepackage[utf8]{inputenc}

\usepackage{microtype}

\usepackage{inconsolata}

\usepackage{graphicx}

\title{Do Audio-Language Models Understand Linguistic Variations?}

\usepackage{fdsymbol}
\newcommand\blfootnote[1]{%
  \begingroup
  \renewcommand\thefootnote{}\footnote{#1}%
  \addtocounter{footnote}{-1}%
  \endgroup
}

\author{
    Ramaneswaran Selvakumar*$^{1}$,
    Sonal Kumar*$^{1}$,
    Hemant Kumar Giri*$^{2}$, \\
    \bf Nishit Anand$^{1}$,
    Ashish Seth$^{1}$,
    Sreyan Ghosh$^{1}$,
    \bf Dinesh Manocha$^{1}$ \\
    $^{1}$University of Maryland, College Park, $^{2}$NVIDIA, Bangalore\\
    \texttt{\{ramans, sonalkum, nishit, aseth125, sreyang, dmanocha\}@umd.edu} \\ 
    \texttt{hgiri@nvidia.com}
}

\usepackage[skins]{tcolorbox}

\newtcolorbox{mybox}[1]{enhanced,sharp corners=all,colback=white,colframe=gray,toprule=0pt,bottomrule=0pt,leftrule=1pt,rightrule=1pt,overlay={
            \draw[gray,line width=1pt] (frame.north west) -- ++(2cm,0pt);
            \draw[gray,line width=1pt] (frame.south east) -- ++(-2cm,0pt);
    },
    coltitle=black,colbacktitle=white,titlerule=0pt,
    title={\vskip5pt\bfseries#1}
}

\begin{document}
\maketitle
\begin{abstract}
Open-vocabulary audio language models (ALMs), like Contrastive Language Audio Pretraining (CLAP), represent a promising new paradigm for audio-text retrieval using natural language queries. In this paper, for the first time, we perform controlled experiments on various benchmarks to show that existing ALMs struggle to generalize to linguistic variations in textual queries. To address this issue, we propose RobustCLAP, a novel and compute-efficient technique to learn audio-language representations agnostic to linguistic variations. Specifically, we reformulate the contrastive loss used in CLAP architectures by introducing a multi-view contrastive learning objective, where paraphrases are treated as different views of the same audio scene and use this for training. Our proposed approach improves the text-to-audio retrieval performance of CLAP by 0.8\%-13\% across benchmarks and enhances robustness to linguistic variation. We make our code publicly available~\footnote{\url{https://github.com/ramaneswaran/linguistic_robust_clap}}\blfootnote{$^*$These authors contributed equally to this work.} 
\looseness=-1
\end{abstract}

\section{Introduction}

\begin{figure}
    \centering
    \includegraphics[width=\columnwidth]{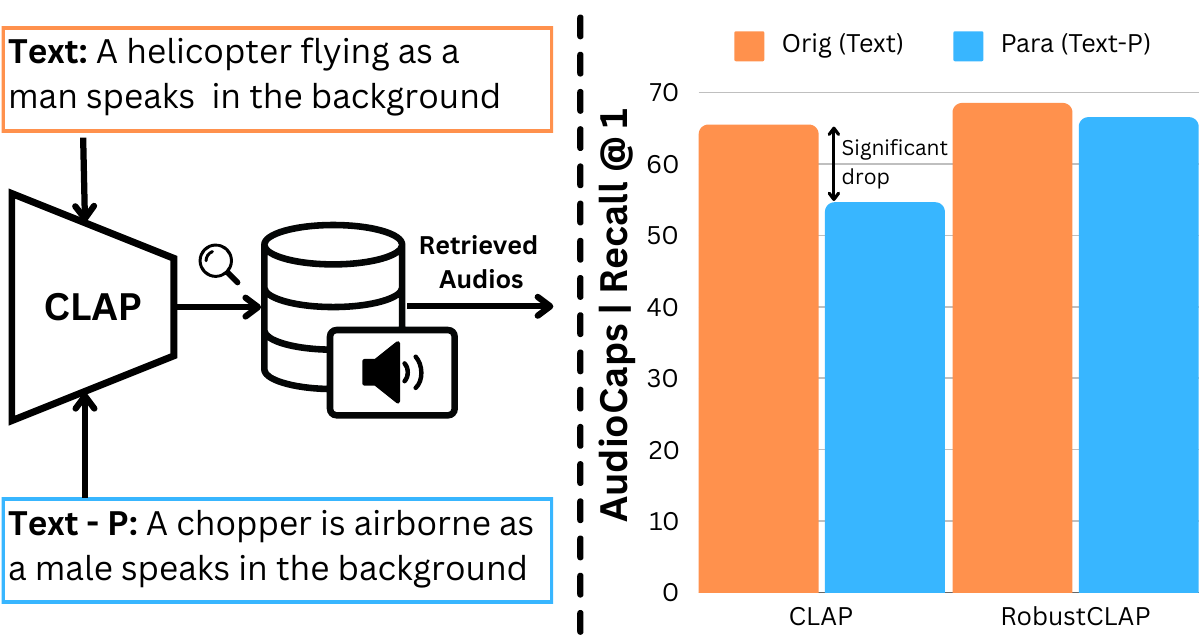}
    \caption{ALMs like CLAP struggle with linguistic variations in queries~(\textbf{Text}), such as paraphrases~(\textbf{Text-P}), resulting in a significant drop in retrieval performance. Our method, RobustCLAP, mitigates this issue while improving overall retrieval accuracy.}
    \label{fig:enter-label}
    \vspace{-2mm}
\end{figure}

As user-generated audio content expands at an unprecedented pace, developing methods to index and search effectively across an ever-growing database becomes crucial. Open-vocabulary audio language models (ALMs) such as CLAP \cite{CLAP2022, CLAP2023} have emerged as a promising solution to this problem, achieving state-of-the-art (SOTA) results in text-based audio retrieval \cite{laionclap2023}. In a typical setting, a user would use a natural language query to describe an acoustic scene with various audio events and then use it to retrieve audio files that match the query. Natural language offers a powerful and intuitive interface for indexing and searching through audio databases. It allows end-users to describe virtually any concept and provides the creative freedom to use linguistically diverse expressions to describe the scene. However, while humans naturally adapt to such linguistic variations, whether ALMs can generalize to these variations at test time remains to be determined. Our preliminary results suggest that the answer is \textbf{\textit{no}}, and ALMs can observe up to a 16\% drop in text-to-audio (T2A) retrieval performance on standard benchmarks with only slight changes in the wording of the text. This limitation can further lead to inconsistent retrieval results across natural language queries with the same intent (see Figure~\ref{fig:enter-label} for an example)  
\looseness=-1

\noindent \textbf{Main Contributions:} To this end, in this paper, we present two novel contributions: 

\begin{enumerate}
\setlength\parskip{0em}
    \item We present the first study to evaluate the robustness of ALMs for T2A retrieval. We construct five new benchmarks (synthetically with human-in-the-loop) to evaluate the performance of CLAP models in T2A retrieval across linguistically varied queries with similar intent. Our evaluation shows a consistent drop in retrieval recall scores (0.1\% - 16\%) across our benchmarks, highlighting the vulnerability to linguistic variation.

    \item We propose RobustCLAP, a simple yet effective method to train CLAP-like ALMs that are robust to linguistic variations in input queries. We continually fine-tune pre-trained CLAP using a novel multi-view contrastive objective that gradually aligns the paraphrased captions with original captions and audio. By training on only a fraction of the original pre-training data, our method improves T2A retrieval performance on the original and paraphrased benchmarks by 0.8\%-13\%, demonstrating increased robustness to linguistic variation while maintaining computational and data efficiency. 
\end{enumerate}

\looseness=-1

\begin{table*}[htbp]
\centering
\setlength{\tabcolsep}{3pt}
\renewcommand{\arraystretch}{1.1}
\resizebox{\textwidth}{!}{
\begin{tabular}{l*{10}{c}}
\toprule
\textbf{Benchmark}~$\longrightarrow$ & \multicolumn{2}{c}{\textbf{AudioCaps}} & \multicolumn{2}{c}{\textbf{Clotho}} & \multicolumn{2}{c}{\textbf{Audioset SL}} & \multicolumn{2}{c}{\textbf{SoundDesc}} & \multicolumn{2}{c}{\textbf{DCASE}} \\
\cmidrule(r){2-3} \cmidrule(r){4-5} \cmidrule(r){6-7} \cmidrule(r){8-9} \cmidrule(r){10-11}
\textbf{Model}~$\downarrow$ & TEST  & TEST-P & TEST & TEST-P & TEST & TEST-P & TEST & TEST-P & TEST & TEST-P \\
\midrule
ML-ACT & 35.53      & 34.87 & 27.54 & 23.90 & 21.52 & 17.91 & 08.72 & 06.06 & 10.12 & 08.77 \\
MSCLAP-22 & 84.74       & 84.63 & \textbf{86.74} & 43.94 & 27.73 & 23.72 & 14.33 & 11.87 & 39.91 & 30.99 \\
MSCLAP-23 & 80.77      & 77.63 & 51.14 & 42.12 & 55.12 & 39.15 & \textbf{38.27} & \textbf{24.89} & 47.84 & 39.21 \\
CompA & 97.17      & 96.23 & 51.28 & 42.49 & 43.03 & 40.24 & 33.32 & 23.56 & 49.54 & 39.51 \\ \hdashline
% RobustCLAP & \textbf{53.21 / 25.48} & \textbf{48.36 / 25.48} & \textbf{83.16 / 25.48} & \textbf{83.83 / 25.48} & \textbf{57.44 / 25.48} & \textbf{53.64 / 25.48} & 25.48 / 25.48 & 21.54 / 25.48 & \textbf{54.66 / 25.48} & \textbf{50.35 / 25.48} \\
LAION-CLAP & 97.80      & 95.92 & 52.03 & 43.98 & 46.91 & 41.94 & 24.62 & 18.09 & 44.73 & 37.81 \\
RobustCLAP & \textbf{98.64}      & \textbf{98.22} & 57.27 & \textbf{53.47} & \textbf{57.44} & \textbf{53.64} & 25.48 & 21.54 & \textbf{54.66} & \textbf{50.35} \\
\bottomrule
\end{tabular}}
\caption{\small Recall@10 scores (higher is better) for text-to-audio retrieval on the original test set (TEST) and paraphrased test set (TEST-P). All ALMs show a consistent, significant drop in performance on TEST-P. RobustCLAP not only improves overall retrieval performance on TEST but also mitigates the drop in TEST-P. The best scores for each benchmark are highlighted in \textbf{bold}.}
\label{tab:main_result}
\end{table*}

\section{Related Work}
\label{sec:related_work}

\subsection{Retrieval With Query Variations}

Although the effects of query variations for text \cite{query_variation_effect, trec8} or image retrieval \cite{kim2024finetuning} have been explored before,   there have been few attempts to address this issue in audio retrieval tasks such as Clotho~\cite{drossos2019clotho} and AudioCaps~\cite{kim-etal-2019-audiocaps}. Most prior efforts to improve audio language models (ALMs) have focused either on scaling the models~\cite{laionclap2023} or enhancing their reasoning capabilities~\cite{ghosh2024compa}. However, as audio retrieval using ALMs is increasingly being used in tasks like audio captioning and question answering~\cite{kong2024audio_flamingo, ghosh2024recap}, ensuring robustness to linguistic variation is critical to maintaining their effectiveness in real-world applications.
\looseness=-1

\subsection{Synthetic Data For Retrieval}

Synthetic data generation has been widely studied in text-based representation learning and information retrieval. InPars~\cite{bonifacio2022inparsdataaugmentationinformation}, InParsv2~\cite{jeronymo2023inparsv2largelanguagemodels} and Promptagator~\cite{dai2022promptagatorfewshotdenseretrieval} generate synthetic queries from unlabelled documents for language encoder training. DINO~\cite{schick2021generatingdatasetspretrainedlanguage} generates synthetic textual similarity pairs for training cross-encoders while Gecko~\cite{lee2024geckoversatiletextembeddings} extensively uses LLMs to generate synthetic queries and hard-negatives. LARMOR~\cite{Khramtsova_2024} uses LLMs to generate synthetic data to adapt textual retrievers to a specific domain. On the other hand, synthetic data for improving audio-language models (ALMs) is still under explored. Approaches like CompA~\cite{ghosh2024compa} pioneer the use of synthetic data to improve general and compositional representation of ALMs and train their models from scratch. In contrast, our approach adapts any off-the-shelf CLAP model and, with minimal additional training, enhances its robustness to linguistic variations while preserving its pre-trained knowledge and capabilities.
\looseness=-1

\section{Methodology}

\subsection{Paraphrased Audio Text Retrieval Benchmark}
\label{sec:paraphrased_benchmarks}
To study the impact of linguistic variation in input queries, we introduce new benchmarks by carefully extending the following five audio-text retrieval benchmarks with their paraphrased captions: 1) AudioCaps~\cite{kim-etal-2019-audiocaps} 2) Clotho~\cite{drossos2019clotho} 3) DCASE~\cite{DCASE2022Workshop} 4) Audioset Strong Labels~\cite{hershey2021benefit_temp_strong} and 5) SoundDesc~\cite{Koepke_2023_benchmark_study}.
\looseness=-1

To obtain the paraphrased captions, we generate new captions such that the vocabulary and the linguistic structure differ while preserving the key concepts and intent. This task requires linguistic expertise and a strong understanding of the concept behind real-world sounds. For instance, accurately differentiating between a bird's "tweet" and a "chirp" involves recognizing subtle differences in tone and context, which are crucial for maintaining the accuracy and relevance of the paraphrases. On the other hand, Large Language Models (LLMs) have shown remarkable aptitude in natural language understanding and real-world common-sense knowledge. Consequently, we propose using LLMs to generate paraphrased captions in a two-step process: Step 1: We instruct the LLM to generate a paraphrase based on custom human-written ICL examples for each benchmark. Step 2: We instruct the LLM to carefully reason~\cite{wei2023chainofthought} whether the paraphrase is accurate and to correct it if required. We detail these steps and give examples below. 
\looseness=-1

\noindent \textbf{Paraphrase Generation:} We instruct the LLM to generate an initial paraphrase (Text-P') of the original caption, such that we describe the acoustic events using varied vocabulary and sentence structures while preserving the original meaning. We give an example below:

\begin{mybox}{Sample caption, paraphrase and corrected paraphrase}
\noindent\textbf{Text:} A person talking which later imitates a couple of meow sounds.

\noindent\textbf{Text-P':} An individual speaks, subsequently mimicking some cat cries.

\noindent\textbf{Text-P:} An individual speaks, subsequently mimicking some cat meows.
\end{mybox}

\noindent \textbf{Paraphrase Correction:} It is crucial that the paraphrased caption accurately conveys the nuances of the original acoustic events. To ensure this, we instruct the LLM to evaluate the paraphrase for both accuracy and specificity, making corrections where necessary. For example, in the paraphrase above, the LLM identified that "cat cries" typically implies a distressed or loud sound, which may not align with the softer or more playful tone often associated with "meows" As a result, the LLM corrects the paraphrase to use "meows" ensuring it better reflects the intended meaning.

For these tasks, we employ LLaMA-3-70B~\cite{llama3modelcard} with in-context learning examples crafted by humans. Following insights from~\cite{shen2022evaluationmetricsparaphrasegeneration} we conducted a qualitative study to evaluate the quality of paraphrase generation and correction. \textbf{Paraphrase Quality:} Human evaluators rated 100 random paraphrases on a 1-5 Likert scale, with an average score of 4.89. \textbf{Paraphrase Correction:} For 50 paraphrases and their corrected versions, evaluators preferred the corrected captions 98\% of the time. We refer readers to Appendix~\ref{appendix:prompts_used},~\ref{appendix:benchmark_paraphrase_evaluation} for additional details on the implementation and evaluation.

\looseness=-1

\subsection{Improving CLAP With Paraphrases}
\label{sec:our_methodology}
To improve the robustness of audio retrieval to linguistic variation, we propose further training of a pre-trained CLAP model using paraphrases of the training data. Specifically, we reformulate the standard CLAP loss as a multi-view contrastive loss that uses two levels of paraphrases as two views to gradually align the text representations with their paraphrased counterparts. At the first level ($T^{\text{p}_1}$), only the linguistic structure is modified while maintaining the same vocabulary. At the second level ($T^{\text{p}_2}$), both the vocabulary and structure are altered. By presenting the model with progressively more complex paraphrases at each training step, we enable it to learn a more generalizable mapping between semantic content and its diverse linguistic expressions. This enhances the model's robustness to linguistic variations in real-world queries.

 A CLAP model takes in an input of an audio-text pair $(A, T)$ and comprises i) audio-encoder $e_A = E(A)$ and (ii) text encoder $e_T = E(T)$. In this notation, we compute similarity score as:
\looseness=-1

\begin{equation}
S(T, I) = \exp \left( \frac{1}{\tau} \cdot \frac{e_T^\top e_A}{\|e_T\| \|e_A\|} \right),
\end{equation}
where $\tau$ is a learned temperature parameter.

\noindent \textbf{Contrastive Loss} For a generated paraphrase $T_{i}^{\text{p}_k}, k \in \{1, 2\}$  produced from the text $T_i$ corresponding to audio $A_i$, we compute the constrastive loss $L^{\text{p}_k}$ as a combination of the following two losses:
\looseness=-1

\begin{equation}
L^{T}_{\text{p}_k} = \sum_i \left[-\log \left( \frac{S(T^{\text{p}_k}_i, T_i)}{\sum_j S(T^{\text{p}_k}_i, T_j)} \right)\right]
\end{equation}
\begin{equation}
L^{\text{A}}_{\text{p}_k} = \sum_i \left[-\log \left( \frac{S(T^{\text{p}_k}_i, A_i)}{\sum_j S(T^{\text{p}_k}_i, A_j)} \right)\right]
\end{equation}

Overall, the final loss is computed as follows: 

\begin{equation}
L_{final} = L_{\text{clap}} + L^{\text{p}_1} + L^{\text{p}_2}
\end{equation}

Here, $L_{\text{clap}}$ is the original CLIP-loss \cite{radford2021learning_clip} used to train the CLAP models. This is necessary to prevent the CLAP model from forgetting its knowledge acquired during pretraining.
\looseness=-1

\section{Experimental Setup}

\noindent \textbf{Training Dataset:} We train our model on a combination of AudioCaps \cite{kim-etal-2019-audiocaps} and Clotho \cite{drossos2019clotho}, which we augment with our two levels of paraphrased captions. 

\noindent \textbf{Evaluation Dataset:} For T-A retrieval, we adopt the evaluation setup from previous work \cite{Koepke_2023_benchmark_study} and employ AudioCaps, Clotho, Audioset SL~\cite{hershey2021benefit_temp_strong}, SoundDesc~\cite{Koepke_2023_benchmark_study} and DCASE~\cite{DCASE2022Workshop}. We evaluate for Recall@10.
\looseness=-1

\noindent \textbf{Baselines:} For baselines we use ML-ACT~\cite{Mei2022metric}, MSCLAP-22~\cite{CLAP2022}, MSCLAP-23~\cite{CLAP2023}, CompA~\cite{ghosh2024compa} and LAION-CLAP \cite{laionclap2023}. We use LAION-CLAP as the base model for RobustCLAP. 
\looseness=-1

\section{Results And Analysis}
\label{sec:results}
\noindent \textbf{Quantitative Analysis:} Table~\ref{tab:main_result} shows that current ALMs struggle with linguistic variations, as evidenced by a significant drop (0.1\%-16\%) in recall scores for paraphrased captions compared to the original captions. In contrast, RobustCLAP not only 1) improves recall scores on the original benchmarks by 0.8\% to 13\% compared to its base model but also 2) mitigates the performance drop on the paraphrased benchmarks, improving scores by 2\% to 12\% compared to the respective second best-performing model. CompA and MSCLAP-23, being trained on SoundDesc, perform better on that dataset. However, they show a significant 10-14\% drop on the paraphrased SoundDesc benchmark, illustrating that fine-tuning can worsen the issue. We evaluate RobustCLAP on zero-shot audio classification tasks using ESC-50~\cite{esc50} and FSD50K~\cite{fsd50k}. CLAP gets a mAP@10 score of 94.25 and 52.20, while RobustCLAP gets 94.07 and 52.81, respectively, on ESC-50 and FSD-50K.  We observe a negligible drop in performance, which indicates that prior knowledge is retained. RobustCLAP also outperforms ALMs on paraphrased audio-to-text retrieval, these results are indicated in Table~\ref{tab:main_result_both_retrieval}.
\looseness=-1

\noindent \textbf{Qualitative Analysis:} We conducted a qualitative experiment to assess how often CLAP retrieves incorrect audio compared to RobustCLAP. We sampled 100 instances where CLAP failed to retrieve the correct audio for a paraphrased query, while RobustCLAP succeeded. We then asked human evaluators to listen to the audio retrieved by CLAP and judge whether they matched the query. The results showed that in 97\% of cases, the retrieved audios were indeed incorrect, while only 3\% were correct. The latter result highlights a challenge inherent in retrieval benchmarks like AudioCaps and Clotho, where a small set of audio files may contain the same acoustic events, mainly when only one or two events are present. 
\noindent Moreover, we observed the following three common mistake patterns. First, CLAP often prioritizes sound events mentioned directly in the query, showing a spurious correlation to non-paraphrased sound events. Second, while the model captures the dominant context or setting of the scene, it frequently lacks precision in identifying all the sound events mentioned in the query. Finally, CLAP fails to recognize attributes or modifiers of a sound event.

\noindent \textbf{Impact Of Sound Event And Attributes:} In an acoustic scene, such as the "steady humming of an engine," the sound event refers to the sound and entity producing the sound (e.g., the "humming of an engine"); sound attributes describe its qualities (e.g., "steady"). We study how paraphrasing these elements affects retrieval performance by instructing the LLM to replace specifically the event and attributes with synonyms while maintaining the original linguistic structure.  In Table~\ref{tab:ablation_result}  we observe that paraphrasing sound attributes leads to a 3.8\% drop in Recall@1, while RobustCLAP significantly reduces this decline to just 0.4\%. However, it is important to note that only 20\% of the samples contain sound attributes, which limits the overall effect of this variation. Paraphrasing sound sources, on the other hand, has a much more significant impact, with recall dropping by as much as 15\%. RobustCLAP mitigates this effect substantially, reducing the performance drop to 3\%.
\looseness=-1

\begin{table}
\centering
\begin{tabular}{@{}ccc@{}}
\toprule
\multirow{2}{*}{\textbf{Dataset}} & \multicolumn{2}{c}{\textbf{Model}}  \\ \cmidrule(l){2-3} 
                         & \textbf{CLAP} & \textbf{RobustCLAP} \\ \midrule
AudioCaps                & 65.51         & 68.54                \\
+ Sound attributes mod.  & 61.96         & 68.12               \\
+ Sound events mod.      & 50.24         & 65.48               \\ \bottomrule
\end{tabular}
\caption{\small We paraphrase sound attributes (row 1) and sound events (row 2), keeping the linguistic structure fixed, to study their impact on R@1 scores. Sound attributes contribute to the drop, while sound events have a greater impact. RobustCLAP mitigates the effects of these paraphrases.}
\label{tab:ablation_result}
\end{table}

\section{Conclusion}

This paper shows that current audio language models lack robustness to linguistic variation in natural language inputs. To demonstrate this phenomenon, we extend several audio-text retrieval benchmarks with paraphrased captions generated through a two-step LLM-based process. To address this issue, we propose a simple mitigation strategy, training CLAP models with a multi-view contrastive loss on a small set of paraphrased data. The resulting model, RobustCLAP, improves retrieval recall scores on the original benchmarks and their paraphrased versions while retaining its prior pre-trained knowledge. We hope our work fuels further studies into improving the robustness of audio-language models.
\looseness=-1

\section{Acknowledgements}

This project is supported in part by NSF\#1910940.

\section{Limitations and Future Work}
As part of future work, we would like to address the following limitations of RobustCLAP:
\begin{itemize}
    \item We utilize LLM to generate the paraphrases for training and testing. Even though we use diverse human-written in-context examples and a correction mechanism, some paraphrases might not be exactly accurate due to hallucinations by the LLM.
    \item We have primarily experimented with diverse audio benchmarks, in future this work can be extended to related domains like music retrieval and speech retrieval. 
    \item We use relatively shorter audio segments, in future this work can extended to long audio.
\end{itemize}

\bibliography{acl_latex}
% \newpage
\appendix
\label{sec:appendix}

\section{Appendix}

In the appendix we provide:

\begin{itemize}
    \item Section~\ref{appendix:dataset_details}: Dataset Details
    \item Section~\ref{appendix:model_details}: Model Details
    \item Section~\ref{appendix:additional_results}: Additional Results
    \item Section~\ref{appendix:additional_implementation_details}: Additional Implementation Details
    \item Section~\ref{appendix:prompts_used}: Prompts Used
\end{itemize}

\section{Dataset Details}
\label{appendix:dataset_details}

In this section we describe in detail the benchmarks that we used for evaluation. In Section~\ref{sec:benchmark_datasets} we describe in detail the datasets that we used. Detailed information about samples present in these are given in Table~\ref{tab:dataset_overview}. In Section~\ref{sec:para_corr_details} we further detail our paraphrase generation and correction mechanism prompts.

\subsection{Benchmark Datasets}
\label{sec:benchmark_datasets}
\noindent\textbf{AudioCaps:}
The Audioscaps \cite{kim-etal-2019-audiocaps} dataset is a large-scale captioning dataset developed by google. It has 46K audio clips with 10 sec duration sourced from Audioset \cite{audioset} along with textual descriptions written by human annotators. They give detailed descriptions of the audio, highlighting specific sound events, their sources, and the context.
\newline
\noindent\textbf{Clotho:}
The Clotho \cite{drossos2019clotho} dataset is an audio captioning dataset with sound clips (15-30 seconds) sourced from Freesound. People have captioned them, describing environments, music, and activities. Along with AudioCaps it is one of the most widely used audio-retrieval benchmarks.
\newline
\noindent\textbf{Audioset SL:}
Audioset SL (Strong Labels) \cite{hershey2021benefit_temp_strong} is a significant component of Google's Audioset project \cite{audioset}, which involves annotating over 2 million 10-second audio clips from YouTube with specific labels. These labels include sounds like "dog barking," "car engine," or "crowd cheering." Although it does not provide full captions, the extensive sound event labeling in Audioset SL provides a rich source for generating artificial captions. We use these temporally strong audio event labels and instruct an LLM to generate a natural language audio caption. 
\newline
\noindent\textbf{SoundDesc:}
The SoundDesc \cite{Koepke_2023_benchmark_study} is a dataset that provides detailed descriptions for diverse sound clips. It includes everyday sounds, natural environments, and specific events. Each clip is paired with a detailed description capturing the sound's essence, source, and context. 
\newline
\noindent\textbf{DCASE:}
The Detection and Classification of Acoustic Scenes and Events (DCASE) \cite{DCASE2022Workshop} dataset is a comprehensive collection of audio recordings. It includes various environments like streets, parks, and indoor settings, each annotated with specific sound event or acoustic scene labels. This dataset is crucial for the DCASE community challenges, fostering advancements in the field. It's essential for developing and evaluating models that recognize and classify sounds in complex environments.

\begin{table}[!ht]
\resizebox{\linewidth}{!}{
\begin{tabular}{@{}lccccccccc@{}}
\toprule
\multicolumn{2}{c}{\textbf{Benchmark}}           & \multicolumn{2}{c}{\textbf{\# Audio Samples}}                 & \multicolumn{2}{c}{\textbf{\# Captions}}                 \\ \midrule
\multirow{2}{*}{\textbf{}}   &                        & \textbf{Train} & \textbf{Test} & \textbf{Train} & \textbf{Test} \\ \cmidrule(l){3-6} 
\multirow{1}{*}{AudioCaps}   &  & 49,275 & 958 & 49,275 & 4,790 \\
\multirow{1}{*}{Clotho}      &  & 3,840 & 1,045 & 19,200 & 5,225 \\
\multirow{1}{*}{Audioset SL} &  & N/A & 1,471 & N/A & 1,471 \\
\multirow{1}{*}{SoundDesc}   &  & 23,085 & 3,250 & 23,085 & 3,250 \\
\multirow{1}{*}{DCASE}       &  & N/A & 997 & N/A & 997 \\ \bottomrule
\end{tabular}}
\caption{\small Overview of the datasets used, including the number of audio samples and captions available for both training and testing.}
\label{tab:dataset_overview}
\end{table}

\begin{table}[ht]
\centering
\renewcommand{\arraystretch}{1.5}
\resizebox{\linewidth}{!}{\begin{tabular}{@{}cl@{}}
\toprule
\multicolumn{1}{l}{\textbf{Score}} & \textbf{Guideline}                                                                   \\ \midrule
1                                  & Completely different meanings with no semantic overlap.                              \\
2                                  & Paraphrased caption shares some similar words but convey different overall meanings. \\
3                                  & Common topic but differ in details or emphasis                                       \\
4                                  & Largely similar meanings with minor variation in detail                              \\
5                                  & The core information being conveyed is same                                          \\ \bottomrule
\end{tabular}}
\caption{\small The likert scale guideline used for paraphrase quality assessment.}
\label{tab:likert_scale}
\end{table}

\subsection{Benchmark Paraphrase Evaluation}
\label{appendix:benchmark_paraphrase_evaluation}
We conduct a qualitative analysis to study both the final paraphrases as well as the performance of paraphrase correction. The volunteers for this study were computer science MS and PhD students.

\noindent\textbf{Paraphrase Generation:} In this experiment, we sample 100 random paraphrases and ask human paraphrases and ask human evaluators to listen to the audio and read the original caption and rate the paraphrase on a LIKERT scale of 1-5. Overall, we obtained an average score of 4.89 indicating that our pipeline of generation and subsequent correction if required is able to generate good paraphrases. The Likert scale guidelines are presented in Table~\ref{tab:likert_scale}.

\noindent\textbf{Paraphrase Correction:} Following \cite{esc50} we conduct an experiment to understand if users prefer the corrected paraphrases as opposed to original paraphrases. We sample 50 random paraphrases and their final corrected versions. We then ask human evaluators to choose one caption which describes the audio better. The corrected versions of the paraphrases were preferred 98\% of the time.

\section{Model Details}
\label{appendix:model_details}
\subsection{Baseline Details}
\label{sec:baselines_details}

\noindent\textbf{ML-ACT}~\citep{Mei2022metric}. This model uses a PANN modlel trained on Audioset~\cite{audioset} and a BERT~\cite{devlin-etal-2019-bert} model and employs the NT-Xent loss adapted from self-supervised learning.

\noindent\textbf{LAION-CLAP}~\citep{laionclap2023}. This is a contrastive language-audio pretraining (CLAP) model from LAION-AI trained on LAION-Audio-630K~\cite{laionclap2023}, a large collection of 633,526 audio-text pairs from different data sources. To improve the model's ability to handle audio inputs of variable lengths and boost overall performance, it integrates a feature fusion mechanism and keyword-to-caption augmentation. This enables the model to effectively align and process both audio and text data for enhanced learning.

\noindent\textbf{LAION-CLAP Music}~\citep{laionclap2023}. This is a music-specific version of the LAION-CLAP model. This version is trained both on audio and music, with the LAION-Audio-630K dataset contributing a major portion of its training data. The details of the music-text data being used for training are not specified.

\noindent\textbf{MS-CLAP 22}~\citep{CLAP2022}. This is a contrastive language-audio pretraining (CLAP) model from Microsoft. This version is trained on 128k audio and text pairs. 

\noindent\textbf{MS-CLAP 23}~\citep{CLAP2023}. This is a follow-up to the MS-CLAP 22, from Microsoft. This version of CLAP uses two innovative encoders and is trained on massive 4.6M audio-text pairs.  To learn audio
representations, the authors trained an audio encoder on 22 audio tasks instead of the standard training of sound event classification. To learn language representations, they trained an autoregressive decoder-only model instead of the standard encoder-only models.

\noindent\textbf{CompA}~\citep{ghosh2024compa}. This is a CLAP model that is trained specifically to enhance its compositional reasoning abilities. The authors introduce improvements to contrastive training by incorporating composition-aware hard negatives, allowing for more precise and focused training. Additionally, they propose a modular contrastive loss designed to help the model learn fine-grained compositional understanding.

% \begin{table}[]
% \centering
% \begin{tabular}{ccc}
% \toprule
% \textbf{Model} & \textbf{ESC-50} & \textbf{FSD-50K} \\ \midrule
% CLAP           & 94.25          & 53.20             \\
% RobustCLAP     & 94.07          & 52.81            \\ \bottomrule
% \end{tabular}
% \caption{\small Zero-shot audio classification results in terms of mAP@10. We observe there is negligible performance decrease for RobustCLAP compared to CLAP}
% \label{tab:zsac_result}
% \end{table}
\begin{table}[ht]
\centering
\renewcommand{\arraystretch}{1.5}
\resizebox{0.7\linewidth}{!}{\begin{tabular}{ccc}
\toprule
\textbf{Model} & \textbf{ESC-50} & \textbf{FSD-50K} \\ \midrule
CLAP           & 94.25          & 53.20             \\
RobustCLAP     & 94.07          & 52.81            \\ \bottomrule
\end{tabular}}
\caption{\small Zero-shot audio classification results in terms of mAP@10. We observe there is negligible performance decrease for RobustCLAP compared to CLAP}
\label{tab:zsac_result}
\end{table}
\section{Additional Results}
\label{appendix:additional_results}
\subsection{Performance On Zero-Shot Audio Classification}

We evaluate CLAP and RobustCLAP on zero-shot audio classification task (ZSAC) on the ESC-50~\cite{esc50} and FSD-50K~\cite{fsd50k} datasets. CLAP gets a mAP@10 score of 94.25 and 52.20, while RobustCLAP gets 94.07 and 52.81, respectively, on ESC-50 and FSD-50K. We observe negligible performance decreases, demonstrating that our approach does not lead to catastrophic forgetting of previously learned knowledge. Fine-tuning the CLAP model on AudioCaps and Clotho enables it to capture the descriptive features (of individual acoustic events), which are beneficial for audio retrieval based on rich natural language descriptions. However, it doesn't necessarily help CLAP learn the discriminative features necessary for zero-shot audio classification. 

\subsection{Error Analysis}
\label{app:error_analysis}

We conduct a manual study of CLAP and RobustCLAP model performance. We sample 100 instances, where for a given paraphrased query, CLAP is not able to correctly retrieve audio whereas RobustCLAP is able to retrieve the audio correctly. We asked human evaluators to listen to the retrieved audio and score whether the audio retrieved by CLAP was correct. The main findings are
\begin{itemize}
    \item In 97\% of the cases, the audio retrieved by CLAP were actually wrong (We highlight some common mistake patterns later in our discussion)
    \item In 3\% of the cases, the audio retrieved was correct according to the given query. This is a challenge inherent in retrieval benchmarks like AudioCaps, Clotho, where a small number of audio files might contain the exact same acoustic events, especially when only one or two events are present.
\end{itemize}

Overall, we were able to verify that CLAP was indeed retrieving the incorrect audio files, whereas RobustCLAP was able to retrieve the correct audio. We noticed some common mistake patterns that we highlight below.

\noindent \textbf{1) Spurious correlation to non-paraphrased sound events:} CLAP tends to prioritize sound events that are directly mentioned in the query without any paraphrasing. In this case, audios which are retrieved may contain an exact sound event such as “wind noise” or “background music” but overall have a completely different meaning compared to the given query. In the examples below the events “background music” and “gurgling and bubbling noises” are spuriously correlated during retrieval
\\
\begin{mybox}{Error Example}
\noindent \textbf{Paraphrased Query:} A man's voice is heard alongside background music and TV noise, then interrupted by kids' giggles and chatter.

\noindent \textbf{Retrieved Audio Description:} A kid is speaking while rattling and tapping sounds are heard amidst the background music, with occasional breathing sounds and mechanisms in the background.

\noindent \textbf{Paraphrased Query:} Continuous music is accompanied by two instances of gurgling and bubbling noises.

\noindent \textbf{Retrieved Audio Description:} Water is poured, splashing and splattering, followed by gurgling and bubbling sounds, with a person breathing in the background towards the end.
\end{mybox}

\noindent \textbf{2) Captures the dominant context but lacks precision:} In this case the model understand the dominant context or the setting of the scene, but fails to precisely capture all the sound events in the query.

\begin{mybox}{Error Example}
\noindent \textbf{Paraphrased Query:} In an urban environment, a man talks as machines and vehicles hum in the background, punctuated by a final thud.

\noindent \textbf{Retrieved Audio Description:} A man is speaking amidst urban traffic noise, accompanied by birds chirping and wind blowing.

\noindent \textbf{Explanation:} CLAP is able to capture the context of a man speaking in urban setting, but does not capture the vehicle hum, but includes bird and wind sound.
\end{mybox}

\noindent \textbf{3) Does not capture sound attributes:} In this case, the CLAP model fails to accurately capture the attributes that act as modifiers to a sound event.

\begin{mybox}{Error Example}
\noindent \textbf{Paraphrased Query:} A serene ambiance is created by an orchestra of bird melodies, punctuated by turkey calls and faint vehicle hums.

\noindent \textbf{Retrieved Audio Description:} A bird is singing along with occasional squawks amidst a constant vehicle noise. 

\noindent \textbf{Explanation:} While CLAP model is able to capture most of the events, listening to the audio shows that a faint vehicle noise (which is in the background and muted) is a big contrast from a constant vehicle noise (which is in the foreground and loud)
\end{mybox}

\subsection{Statistical Significance Test}

We use a bootstrapping method to collect recall metrics for both CLAP and RobustCLAP. This involves repeatedly sampling with replacement from the test set and then computing the recall for each resampled set. These sets of recall values are used to perform a t-test, and we conclude that the improvement of RobustCLAP over CLAP is statistically significant.

\section{Additional Implementation Details}
\label{appendix:additional_implementation_details}

\begin{table}[ht]
\centering
\renewcommand{\arraystretch}{1.5}
\resizebox{\linewidth}{!}{\begin{tabular}{@{}lcl@{}}
\toprule
\multicolumn{1}{c}{\textbf{Model}} & \textbf{\# Params} & \multicolumn{1}{c}{\textbf{Link}}                  \\ \midrule
ML-ACT                             & 140M                & \url{https://github.com/XinhaoMei/audio-text\_retrieval} \\
MSCLAP22                           & 196M                & \url{https://github.com/microsoft/CLAP}                  \\
MSCLAP23                           & 159M                & \url{https://github.com/microsoft/CLAP}                  \\
CompA                              & 300M                & \url{https://github.com/Sreyan88/CompA}                  \\
LAION-CLAP                         & 158M               & \url{https://github.com/LAION-AI/CLAP/}                  \\ \bottomrule
\end{tabular}}
\caption{\small ALMs used in our project and their size (in millions of parameters). We use official implementations of these models}
\label{tab:model_parameters}
\end{table}

\subsection{Model Parameters}

The ALMs that consists of an audio-encoder and a BERT like text encoder. Typically these models are under 300M parameters, refer to Table~\ref{tab:model_parameters} for more details. We use a Llama3-70B which consists of 70B parameters to generate paraphrases for training and validation. 

\subsection{Compute Infrastructure}

RobustCLAP is trained on four NVIDIA A100 GPUs and takes around 2 hours to converge. Inference only requires 1 A100 GPU. To perform inference on the LLama3-70B model we use 4 NVIDIA A100 GPUs. 

\subsection{Implementation Software And Packages:}

For all the ALMs that we implement we use their original GitHub repository. We provide links to these in Table~\ref{tab:model_parameters}. We build RobustCLAP on top of LAION-CLAP repository and use their base models. To perform accelerated inference on Llama3-70B we use vllm~\footnote{\url{https://github.com/vllm-project/vllm}}

\subsection{Potential Risks:}

Our approach involves using an LLM to generate paraphrases for training and evaluation. While LLMs can sometimes hallucinate or produce incorrect or toxic outputs, we mitigated these risks through a qualitative analysis of the generated paraphrases. In our analysis, we observed no toxic outputs, and the paraphrases were of consistently high quality.

\section{Prompt Details}
\label{appendix:prompts_used}

\onecolumn\subsection{Paraphrase Generation And Correction Prompts}
{ % Start a group for local changes

\renewcommand{\familydefault}{\ttdefault} % Change to typewriter font
\normalfont % Apply the change
\begin{flushleft}
\textbf{Paraphrase Generation Prompt}\newline\\
<s>[INST] I will provide you with an audio caption of an audio. Paraphrase the caption while accurately describing the nuances and technical terms. Here are some input-output examples:\\
Input Caption: Gunfire, followed by a click and shattering glass. \\
Paraphrase Caption: Shots ring out, then a click and glass breaks into fragments.\newline\\

Input Caption: Pots clatter as water flows from a turned-on faucet. \\
Paraphrase Caption: Utensils clatter while liquid streams from an open tap.\newline\\

Input Caption: A man and woman laugh, followed by a man shouting and a woman joining in with childlike giggles. \\
Paraphrase Caption: A couple chuckles, then a male yells, and a female responds with youthful giggles.\newline\\

Input Caption: A woman delivers a formal address. \\
Paraphrase Caption: A female presents an official speech.\newline\\

Input Caption: High-pitched snoring echoes repeatedly. \\
Paraphrase Caption: Sharp snores resound over and over.\\
Here is the Input Caption: Constant rattling noise and sharp vibrations [/INST]
\newline\\

\textbf{Prompt Paraphrase Correction}\newline\\
<s>[INST] I will provide you with an audio caption of an audio and its paraphrase. I want you to tell me if the caption is accurately paraphrased especially check if the paraphrased sound events convey the same nuance.Suggest if correction is required and provide corrected paraphrase by give your reasoning. Here are some input-output examples:\\
Input Caption: :A man talking as metal clanks together followed by footsteps on grass while insects buzz in the background. \\
Paraphrase Caption: A male speaks as metallic objects collide, succeeded by the sound of steps on a lawn amidst a gentle humming of bugs.\\
Correction: foo \\
Corrected Paraphrase Caption:A male speaks as metallic objects clatter, succeeded by the sound of steps on a lawn amidst a gentle humming of bugs\\
Reasoning: The term "collide" broadly implies contact but lacks the specific metallic sound detail conveyed by "clank." Using "metallic objects chime" or "metallic clatter" would better capture the resonant sound characteristic of metal without reusing the original word.\newline\\
Input Caption:Men speak as someone snores.\\
Paraphrase Caption: Males converse amidst a person's heavy breathing.\\
Correction: foo\\
Corrected Paraphrase Caption:Males converse amidst a person's disruptive nasal noises.\\
Reasoning: "Heavy breathing" generally suggests deep breaths and lacks the unique, disruptive nature associated with snoring. A phrase like "disruptive nasal noises" more accurately conveys the irritating and unmistakable sounds of snoring, highlighting its potential to interrupt or disturb. This emphasizes not only the sound but also the common reaction to it.\newline\\
Input Caption:An ambulance travels with the siren blaring loudly and moves through traffic.\\
Paraphrase Caption: A rescue vehicle speeds along with its alarm wailing at full volume and navigates through congested roads.\\
Correction: bar\\
Corrected Paraphrase Caption:Not required\\
Reasoning: This is accurate.\newline\\
Input Caption:An idle vehicle engine running.\\
Paraphrase Caption: A stationary car motor hums continuously.\\
Correction: bar.\\
Corrected Paraphrase Caption:Not required.\\
Reasoning: This is accurate.\newline\\
Input Caption:A toy helicopter flying followed by wind blowing into a microphone.\\
Paraphrase Caption: A miniature aircraft whirs as it moves through the air, then a gust of air hits the recording device.\\
Correction: foo\\
Corrected Paraphrase Caption:A miniature aircraft whirs as it moves through the air, followed by wind rushing continuously against the recording device.\\
Reasoning: The phrase "wind blowing into a microphone" suggests a continuous or ambient wind noise, which is not precisely captured by "a gust of air hits the recording device." To better reflect the ongoing nature of the sound, the paraphrase could use "as wind rushes against the recording device" or as 'wind continuously interacts with the recording device.'\newline\\
Input Caption:A man and a woman talking as paper crinkles.\\
Paraphrase Caption: A male and female converse amidst the rustling of documents.\\
Correction: bar\\
Corrected Paraphrase Caption:Not required\\
Reasoning: This is accurate.\newline\\
Input Caption:White noise and then birds chirping.\\
Paraphrase Caption: A gentle hum precedes the sweet sounds of avian creatures.\\
Correction: foo\\
Corrected Paraphrase Caption:A continuous static hum precedes the crisp chirping of birds.\\
Reasoning: The term 'gentle hum' suggests a softer, more subdued sound compared to 'white noise,' which generally implies a more consistent, static-like background noise. To maintain the specific quality of 'white noise,' a more precise description like 'continuous static hum' could be used instead of 'gentle hum.' Additionally, 'the sweet sounds of avian creatures' does not capture the distinctive, rhythmic chirping of birds. A term like 'crisp chirping' would more accurately reflect the clear, melodic nature of bird calls.\newline\\
Input Caption: Music is playing.\\
Paraphrase Caption: A melody fills the air.\\
Correction:[/INST]\newline\\
\end{flushleft}
}
\label{sec:para_corr_details}
\newpage

% % \section{Additional Details: RobustCLAP}

% % \subsection{Baseline Details}

% % \textbf{CLAP}

\onecolumn\subsection{\textsf{Paraphrase Samples}}
{ % Start a group for local changes
\renewcommand{\familydefault}{\ttdefault} % Change to typewriter font
\normalfont % Apply the change
\begin{flushleft}
\textbf{AudioCaps}\\
\underline{TEXT}: People are talking while a motor vehicle engine is revving. \\
\underline{TEXT-P}: A group of individuals engage in conversation amidst a car engine's loud, rapid revving.
\newline\\
\underline{TEXT}: A lady laughing while a baby cries, then the lady speaks and a couple men also talk as well\\
\underline{TEXT-P}: A female bursts into laughter as an infant wails, then she utters words and a pair of males join in the conversation too.
\newline\\
\underline{TEXT}: Clicks followed by gunshots and breathing then some speaking\\
\underline{TEXT-P}: Series of clicks precede gunfire, labored breathing, and subsequent conversation.
\newline\\
\underline{TEXT}: Metal clanking followed by steam hissing as a truck engine is running then accelerating\\
\underline{TEXT-P}: Clattering metal sounds precede a continuous hissing of steam as a lorry's motor hums and gains speed.
\newline\\
\underline{TEXT}: A goat bleating with people speaking\\
\underline{TEXT-P}: A goat lets out a loud, nasal cry while individuals converse.\newline\\
%CLOTHO
\textbf{Clotho}\\
\underline{TEXT}: Water goes down a drain pipe while water is dripping. \\
\underline{TEXT-P}: Liquid flows down a drainage tube as droplets fall.
\newline\\
\underline{TEXT}: The ripping of paper occurs at evenly spaced intervals.\\
\underline{TEXT-P}: The tearing of a document happens at regular time gaps.
\newline\\
\underline{TEXT}: Metal sliding together such as swords or knives.\\
\underline{TEXT-P}: Metallic blades scraping against each other, similar to clashing swords.
\newline\\
\underline{TEXT}: Someone walking slowly, their feet are crunching leaves.\\
\underline{TEXT-P}: A person strolls at a slow pace, their footsteps crushing foliage.
\newline\\
\underline{TEXT}: A man and woman are talking among themselves while others chat in the background.\\
\underline{TEXT-P}: A gentleman and lady converse privately amidst murmurs of surrounding discussions.\newline\\
%Audioset SL
\textbf{Audioset SL}\\
\underline{TEXT}: A camera shutter is snapped twice during an ongoing music session. \\
\underline{TEXT-P}: A camera shutter clicks twice, punctuating the ongoing musical performance. 
\newline\\
\underline{TEXT}: A vehicle is moving through an urban area filled with traffic noise, accompanied by a rooster's crowing and various bird vocalizations. \\
\underline{TEXT-P}: A car navigates through a bustling cityscape with constant traffic din, interspersed with a rooster's loud, shrill crowing and varied bird vocalizations. 
\newline\\
\underline{TEXT}: Music plays while occasional mechanisms and impact sounds are heard, including thuds and a ticking sound, with additional sound effects.\\
\underline{TEXT-P}: Music plays alongside intermittent mechanical noises, occasional thuds, and a steady ticking, accompanied by additional sound effects.
\newline\\
\underline{TEXT}: A vehicle is accelerating in the midst of a noisy crowd and hubbub with people talking in the background. \\
\underline{TEXT-P}: Amidst a chaotic and loud crowd with murmurs of conversation, a vehicle rapidly gains speed.
\newline\\
\underline{TEXT}: A man speaks in a small room filled with mechanisms, where rodents are scurrying around.\\
\underline{TEXT-P}: A male voice is audible amidst machinery sounds and rodents scurrying around in a confined space.\newline\\
%SoundDesc
\textbf{SoundDesc}\\
\underline{TEXT}: A monkey makes close-up snake alarm calls with birds in the background. \\
\underline{TEXT-P}:A monkey's loud, close-up warning cries mix with bird sounds.
\newline\\
\underline{TEXT}: Two seals challenge each other with close-up calls and snorts, accompanied by surf. \\
\underline{TEXT-P}: Two seals get up close and personal, growling and snorting at each other. 
\newline\\
\underline{TEXT}: Chaffinches, crossbills, and great tits sing amidst the rustling of trees in high wind.\\
\underline{TEXT-P}: Birds like chaffinches and crossbills belt out their tunes as the trees creak in the gusty breeze.
\newline\\
\underline{TEXT}: A vintage car approaches, stops, and switches off. \\
\underline{TEXT-P}: Old-school wheels roll up, come to a stop, and kill the engine.
\newline\\
\underline{TEXT}: A lesser black-backed gull vocalizes closely, then attacks a juvenile, amidst herring gulls.\\
\underline{TEXT-P}: A lesser black-backed gull squawks loudly, then swoops in on a young bird, surrounded by herring gulls.\newline\\
%DCASE
\textbf{DCASE}\\
\underline{TEXT}: A continuous chirp while birds chatter quietly in the background and then a meow from a cat. \\
\underline{TEXT-P}:Birds chat softly in the background as a steady chirp flows, interrupted by a cat's meow.
\newline\\
\underline{TEXT}: A truck drives by while a woman speaks in the background. \\
\underline{TEXT-P}: A woman chats away as a truck zooms past in the distance. 
\newline\\
\underline{TEXT}: A train is coming closer and closer, then passes.\\
\underline{TEXT-P}: A locomotive approaches, getting louder, then zooms by.
\newline\\
\underline{TEXT}: Continuous 8-bit arcade game sounds that are building in pitch. \\
\underline{TEXT-P}: Retro arcade sounds amp up, getting higher pitched
\newline\\
\underline{TEXT}: A group of girls laughing harder and louder as time goes by.\\
\underline{TEXT-P}: Girls' giggles escalate to uncontrollable laughter over time.
\end{flushleft}
} % End the group

\begin{table*}[!ht]
\resizebox{\textwidth}{!}{
\begin{tabular}{@{}clcccccccc@{}}
\toprule
\multicolumn{2}{c}{\textbf{Retrieval Type $\longrightarrow$}}           & \multicolumn{4}{c}{\textbf{Text-to-Audio Retrieval}}                 & \multicolumn{4}{c}{\textbf{Audio-to-Text Retrieval}}                 \\ \midrule
\multirow{2}{*}{\textbf{Benchmark}}   & \multirow{2}{*}{\textbf{Model}} & \multicolumn{2}{c}{R@1 $\uparrow$} & \multicolumn{2}{c}{R@10 $\uparrow$} & \multicolumn{2}{c}{R@1 $\uparrow$} & \multicolumn{2}{c}{R@10 $\uparrow$}  \\ \cmidrule(l){3-10} 
                             &                        & TEST & TEST-P & TEST & TEST-P & TEST & TEST-P & TEST & TEST-P   \\ \midrule
\multirow{6}{*}{AudioCaps}   & ML-ACT     & 08.36      & 07.92       & 35.53      & 34.87       & 07.97         & 06.42           & 29.17          & 26.14           \\
                             & MSCLAP-22  & 39.18       & 36.99       & 84.74       & 84.63        & 33.33         & 16.50           & 79.41          & 59.35           \\
                             & MSCLAP-23  & 37.30       & 24.24       & 80.77      & 77.63       & 28.42         & 22.57           & 77.84          & 68.44           \\
                             & CompA      & 67.81       & 58.72       & 97.17      & 96.23       & \textbf{46.02}         & 31.17           & 88.59          & 78.97           \\
                             & LAION-CLAP & 65.51      & 54.64       & 97.80      & 95.92       & 43.36         & 32.60           & \textbf{89.86}          & 80.14           \\
                             & RobustCLAP & \textbf{68.54}      & \textbf{66.35}       & \textbf{98.64}      & \textbf{98.22}       & 45.76         & \textbf{40.96}           & 89.34          & \textbf{86.31}           \\ \midrule
\multirow{6}{*}{Clotho}      & ML-ACT     & 12.87      & 11.42      & 27.54      & 23.90       & 13.20         & 12.03           & 52.71         & 48.87           \\
                             & MSCLAP-22  & 36.19      & 29.78      & \textbf{86.74}       & 43.94       & 19.76         & 12.24           & 51.89          & 45.93           \\
                             & MSCLAP-23  & 37.03      & 30.47      & 51.14      & 42.12        & 22.87         & 16.26           & 61.53          & 51.19           \\
                             & CompA      & 36.39      & 29.11      & 51.28      & 42.49       & 17.14         & 11.97           & 53.44          & 44.34           \\
                             & LAION-CLAP & 36.75      & 32.54      & 52.03      & 43.98       & 37.03         & 30.72           & 81.91          & 74.83           \\
                             & RobustCLAP & \textbf{39.43}      & \textbf{38.66}      & 57.27      & \textbf{53.48}       & \textbf{39.43}         & \textbf{37.32}           & \textbf{82.49}          & \textbf{82.30}           \\ \midrule
\multirow{6}{*}{Audioset SL} & ML-ACT     & 04.31          & 04.01          & 21.52          & 17.91           & 05.54         & 03.77           & 22.02          & 18.91           \\
                             & MSCLAP-22  & 06.45       & 04.74       & 27.73      & 23.72       & 07.00         & 05.57           & 30.38          & 26.03           \\
                             & MSCLAP-23  & 21.02      & 16.85      & 55.12      & 39.15       & 15.43         & 13.46           & 51.66          & 46.29           \\
                             & CompA      & 11.82      & 10.19      & 43.03      & 40.24       & 15.70         & 13.18           & \textbf{53.36}          & 43.77           \\
                             & LAION-CLAP & 14.41      & 11.62      & 46.91      & 41.94       & \textbf{16.52}         & 11.90           & 52.75          & 43.71           \\
                             & RobustCLAP & \textbf{21.82}      & \textbf{19.10}       & \textbf{57.44}      & \textbf{53.64}       & 15.84         & \textbf{14.41}           & 50.37          & \textbf{47.99}           \\ \midrule
\multirow{6}{*}{SoundDesc}   & ML-ACT     & 01.10          & 00.65          & 08.72          & 06.06           & 00.74         & 00.60           & 08.96          & 07.32           \\
                             & MSCLAP-22  & 02.33       & 01.96       & 14.33      & 11.80       & 01.84           & 01.44          & 09.72  & 09.63          \\
                             & MSCLAP-23  & \textbf{09.75}       & \textbf{05.53}       & \textbf{38.27}      & \textbf{24.89}       & \textbf{06.58}         & \textbf{05.72}           & \textbf{26.36}          & \textbf{25.60}           \\
                             & CompA      & 06.80        & 04.03       & 33.32      & 23.56       & 04.21         & 03.32           & 20.86          & 17.26           \\
                             & LAION-CLAP & 05.82       & 03.17       & 24.62      & 18.09       & 03.23         & 02.34           & 17.75          & 13.63           \\
                             & RobustCLAP & 05.45       & 05.02       & 25.48      & 21.54       & 03.78         & 02.95           & 19.08          & 16.92           \\ \midrule
\multirow{6}{*}{DCASE}       & ML-ACT     & 01.47          & 01.12          & 10.12          & 08.77           & 02.93         & 02.87           & 13.50          & 11.63           \\
                             & MSCLAP-22  & 09.82       & 07.02      & 39.91      & 30.99       & 10.53         & 05.71           & 39.71          & 27.08           \\
                             & MSCLAP-23  & 13.84      & 10.43      & 47.84      & 39.21       & 15.64         & 11.73           & 49.24          & 40.92           \\
                             & CompA      & 14.84      & 10.61      & 49.54      & 39.51       & 14.44         & 08.92           & 48.54          & 35.10           \\
                             & LAION-CLAP & 13.34      & 11.23      & 44.73      & 37.81       & \textbf{17.25}         & 10.93           & \textbf{54.86}          & 43.53           \\
                             & RobustCLAP & \textbf{17.45}      & \textbf{15.95}      & \textbf{54.66}      & \textbf{50.35}       & 14.84         & \textbf{13.14}           & 48.65          & \textbf{45.94}           \\ \bottomrule
\end{tabular}}
\caption{\small Text-to-audio and audio-to-text on the original test set (TEST) and paraphrased test set (TEST-P). All ALMs show a consistent, significant drop in performance on TEST-P. RobustCLAP not only improves overall retrieval performance on TEST but also mitigates the drop in TEST-P. The best scores for each benchmark are highlighted in \textbf{bold}.}
\label{tab:main_result_both_retrieval}
\end{table*}

\end{document}